\documentclass[submission,copyright,creativecommons]{eptcs}
\providecommand{\event}{MEMICS 2016} 
\usepackage{breakurl}                
\usepackage{underscore}              

\usepackage{tabularx}
\usepackage{multirow}
\usepackage{siunitx}
\usepackage{dcolumn}
\usepackage{array}
\usepackage[table]{xcolor}
\usepackage{amsmath}
\usepackage[figuresright]{rotating}

\newcommand{\rotatedHeader}[2][l]{\rotatebox{90}{\renewcommand{\arraystretch}{0.75}\begin{tabular}[#1]{@{}l}#2\end{tabular}}}
\newcommand{\cc}{\cellcolor{black!15}}  

\newcommand{\eacirc}{EACirc}
\newcommand{\niststs}{NIST STS}
\newcommand{\dieharder}{Dieharder}
\newcommand{\testu}{TestU01}
\newcommand{\caesar}{CAESAR}

\title{Avalanche Effect in Improperly Initialized\\ \caesar{} Candidates}
\author{Martin Ukrop
\institute{Centre for Research on Cryptography and Security,\\
Faculty of Informatics,\\
Masaryk University, Brno, Czech Republic}
\email{mukrop@mail.muni.cz}
\and
Petr \v{S}venda
\institute{Centre for Research on Cryptography and Security,\\
Faculty of Informatics,\\
Masaryk University, Brno, Czech Republic}
\email{svenda@fi.muni.cz}
}
\def\titlerunning{Avalanche effect in \caesar{}}
\def\authorrunning{M. Ukrop \& P. \v{S}venda}
\begin{document}
\maketitle

\begin{abstract}
Cryptoprimitives rely on thorough theoretical background, but often lack basic usability features making them prone to unintentional misuse by developers. We argue that this is true even for the state-of-the-art designs. Analyzing 52 candidates of the current \caesar{} competition has shown none of them have an avalanche effect in authentication tag strong enough to work properly when partially misconfigured. Although not directly decreasing their security profile, this hints at their security usability being less than perfect.\footnote{Paper details available at \url{crcs.cz/papers/memics2016}}
\end{abstract}

\section{Introduction}

Nowadays, experts realize that having cryptography attack-resistant from the theoretical point of view is not sufficient since many attacks are caused by improper use of otherwise sound cryptographic primitives. Developers routinely produce horrendous implementations (at least from the point of security) when they neglect to properly set initialization vectors or ignore the requirement of unique sequence numbers. Recently, Cairns and Steel outlined their vision for developer-resistant cryptography~\cite{dev-resistant-crypto} with designs that cannot be misused by the programmer.

The question the security-optimist would ask is: \textit{Is that not the case only for old primitives, old protocols and old designs? Are new designs also prone to developer misuse?} We argue the problem is still open -- we tested 52 participants of the current state-of-the-art cryptographic competition by checking the avalanche effect of the candidates in settings simulating partial misconfiguration.

It is long known that cryptographic primitives such as ciphers, hash functions and message authentication codes should produce seemingly random outputs. Further requirements ask for outputs to change unpredictably with respect to variations in the input. The strict avalanche criterion, as introduced by Webster and Tavares in 1985~\cite{sac}, is one way to formalize this. It is satisfied if, whenever a single input bit is complemented, each of the output bits changes with a 50\% probability. It is commonly used for assessing the security of hash functions, though using it as a randomness test has also been done before~\cite{sac-randomness-test}.

\begin{figure}[t]
\centering
\includegraphics[width=\textwidth]{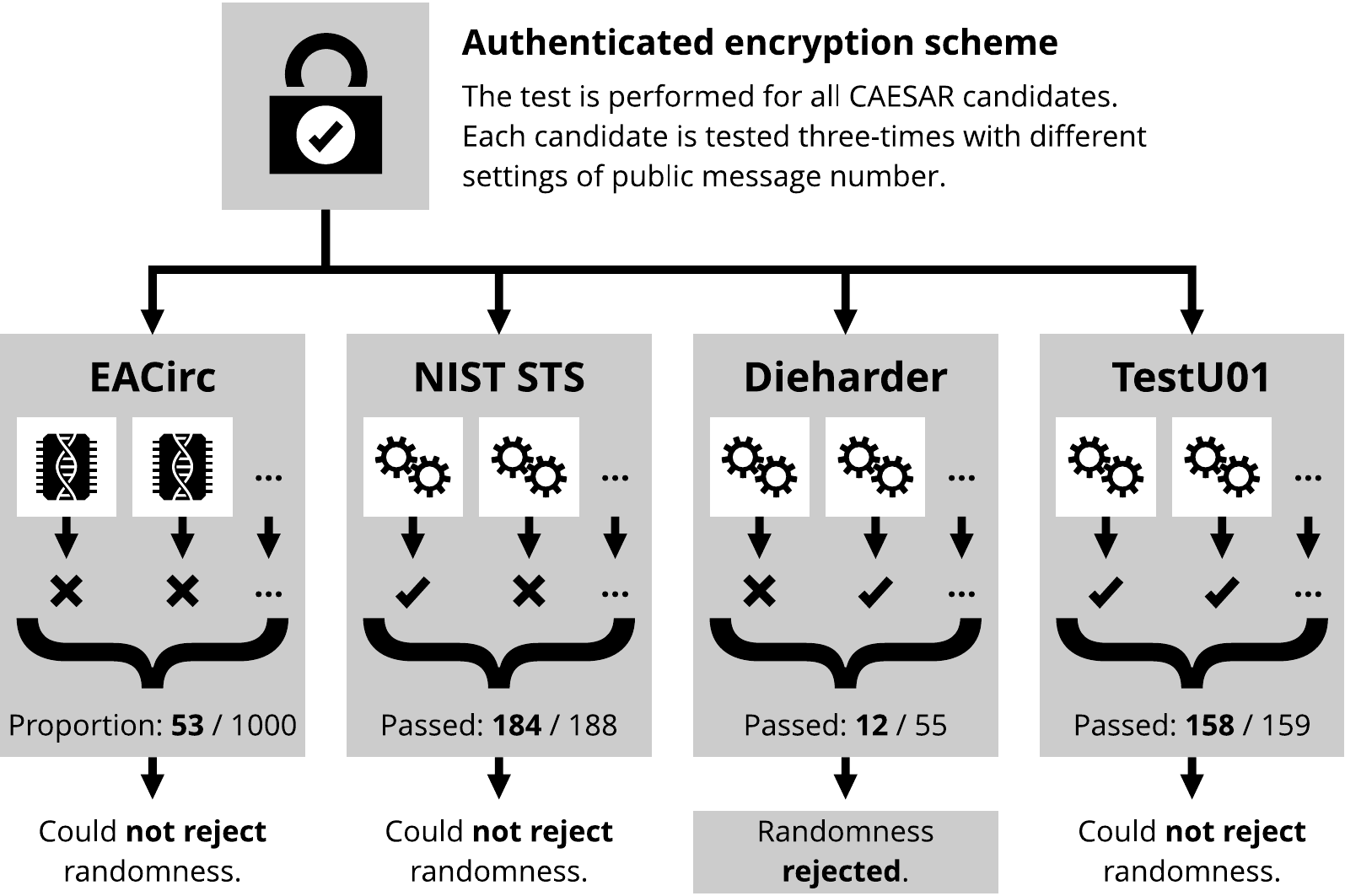}
\caption{High-level overview of the performed experiments. Firstly, a \caesar{} cipher is used to generate a stream of authentication tags. The randomness of this stream is then assessed by 4 tools (\eacirc{} and 3 statistical testing suites). If the design is good, it should exhibit an avalanche effect strong enough for the stream to look random.}
\label{fig:experiment-overview}
\end{figure}

In this paper, we scrutinize submissions of the ongoing \caesar{} competition (\textit{Competition for Authenticated Encryption: Security, Applicability, and Robustness})~\cite{caesar-competition}. The authentication tags produced by all candidates are examined using four different software tools: three standard statistical batteries (\niststs{}~\cite{nist-sts-documentation}, \dieharder{}~\cite{dieharder} and \testu{}~\cite{testu01}) and a novel genetically-inspired framework (\eacirc{}~\cite{ccis2014}). The overview of the main experiment idea is depicted in Figure~\ref{fig:experiment-overview}.

The analysis was done separately for three different settings of the public message number (fixing it to zero, using a counter and generating unique random value each time). It turned out that none of the tested \caesar{} candidates had an avalanche effect strong enough to produce random-looking tags in the most seriously misconfigured case with zero public message numbers (thus avalanching from only a very few changed bits in plaintext). In the case of counter-based and random public message numbers, the ciphers fared much better.

Firstly, in Section~\ref{sec:related-research}, the paper gives an overview of the related research. Then the basics of authenticated encryption are explained along with the essentials of \caesar{} competition (Section~\ref{sec:authenticated-encryption}). The following sections summarize the way of generating the tested data (Section~\ref{sec:tested-data-streams}) and the tools used for the analysis (Section~\ref{sec:randomness-testing-tools}). Lastly, the results and their interpretation are given in Section~\ref{sec:results}.

\section{Related research}
\label{sec:related-research}

As \caesar{} competition is an on-going initiative with many submissions, there are still not many publications thoroughly examining the security of all the proposed algorithms. F.~Abed et al.~\cite{caesar-classification} give an excellent overview of the candidates along with a classification with regard to their core primitives. K.~Hakju and K.~Kwangjo~\cite{caesar-survival} discuss the features of authenticated encryption and predict the essential characteristics of the submissions to survive the \caesar{} competition.

Probably the most comprehensive competition-wide analysis so far has been done by M.~Saarinen~\cite{brutus} using the BRUTUS automatic cryptanalytic framework. Deeper analysis exists only on a per-candidate basis. For example, R.~Ankele in his Ph.D.\ thesis~\cite{caesar-thesis} analyses the \textit{COPA} authenticated encryption composition scheme used in several \caesar{} candidates. M.~Nandi in his 2014 paper~\cite{caesar-cobra-poet} demonstrates a forging attack on \textit{COBRA} and \textit{POET} ciphers.

Numerous works tackled the problem of assessing randomness of outputs from other cryptoprimitives. E.\ Simion~\cite{ieee-simion2015} gave a nice overview of statistical requirements for cryptographic primitives in his work. The Ph.D.\ thesis of K.\ Jakobsson~\cite{jakobsson} provides both a good theoretical background and a comparison of commonly available tools for random number testing. Its results are based on assessing a variety of pseudo-random and quantum random number generators.

Cryptographic competitions are often the target of these analyses since the unified function API allows for easy evaluation of a high number of schemes. M.\ Turan et al.~\cite{estream-statistical-testing} performed a detailed examination of eStream phase 2 candidates (both full and reduced-round) with \niststs{} and structural randomness tests, finding six ciphers deviating from expected values. In 2010, Doganaksoy et al.~\cite{sha3-statistical-testing} applied the same battery, but only a subset of tests to SHA-3 candidates with a reduced number of rounds as well as only to their compression functions.

A different strategy is employed in the \eacirc{} framework -- it uses a genetically-inspired process to find a successful distinguisher (function capable for differencing between cipher output and random stream). The framework has been used for assessing the randomness of outputs produced by the round-limited eSTREAM and SHA-3 candidates~\cite{ccis2014,secrypt2014}. Although still falling behind in some cases, this approach surpasses \niststs{} in a few instances.

\section{Authenticated encryption}
\label{sec:authenticated-encryption}

A cryptosystem for authenticated encryption simultaneously provides confidentiality, integrity, and authenticity assurances on data -- decryption is combined in a single step with integrity verification. Authenticated ciphers are often built as various combinations of block ciphers, stream ciphers, message authentication codes, and hash functions. There are many examples commonly used today, such as the \textit{Galois/counter mode} (GCM)~\cite{gcm} based on block ciphers.

Combining privacy and integrity assurances into a single scheme has tremendous advantages as combining a confidentiality mode with an authentication mode could be error prone and difficult\footnote{\textit{``It is very easy to accidentally combine secure encryption schemes with secure MACs and still get insecure authenticated encryption schemes."}~\cite{ae-quote}}. Therefore, following a long tradition of cryptography competitions, \caesar{}~\cite{caesar-competition} aims to create a portfolio of authenticated encryption systems intended for wide public adoption.

Each submission in \caesar{} specifies a family of authenticated ciphers. Family members differ only in parameters (e.g.\ key length, the number of internal rounds). There were 56 different designs submitted to the first round. Taking into account all possible parameter sets, this amounts to 172 independent schemes. Till the announcement of the second-round candidates, nine ciphers were withdrawn by their authors. On July 7\textsuperscript{th} 2015, 29 ciphers were chosen for the second round. Later, on August 15\textsuperscript{th} 2016, 15 ciphers out of these were selected for the third round.

Our goal was to test as many authenticated encryption schemes as possible. Using \caesar{} candidates enabled us to test many ciphers and many configurations automatically due to the shared API. All the candidate source codes were taken from the 1\textsuperscript{st}-round SUPERCOP repository managed by eBACS~\cite{supercop}.

In the end, there were 168 different ciphers tested in all performed experiments. From 172 submitted independent schemes (56 designs with various parameter sets), six were not tested. Firstly, we could not get the \textit{AVALANCHE} candidates working properly (segmentation fault while running). Secondly, \textit{Julius} did not compile due to problems with the inclusion of the external AES routines provider. Thirdly, \textit{POLAWIS} seemed not to have followed the prescribed API. Lastly, the implementation of \textit{PAES} is probably faulty, since it did not pass our encrypt-decrypt sanity test. We might have been able to fix most of these cases, but doing so would require extensive interventions in the code increasing the possibility of error. Apart from the submitted candidates, we tested two versions of \textit{AES/GCM} referenced by the \caesar{} committee as a design baseline.

\section{Tested data streams}
\label{sec:tested-data-streams}

\begin{figure}[t]
\centering
\includegraphics[width=\textwidth]{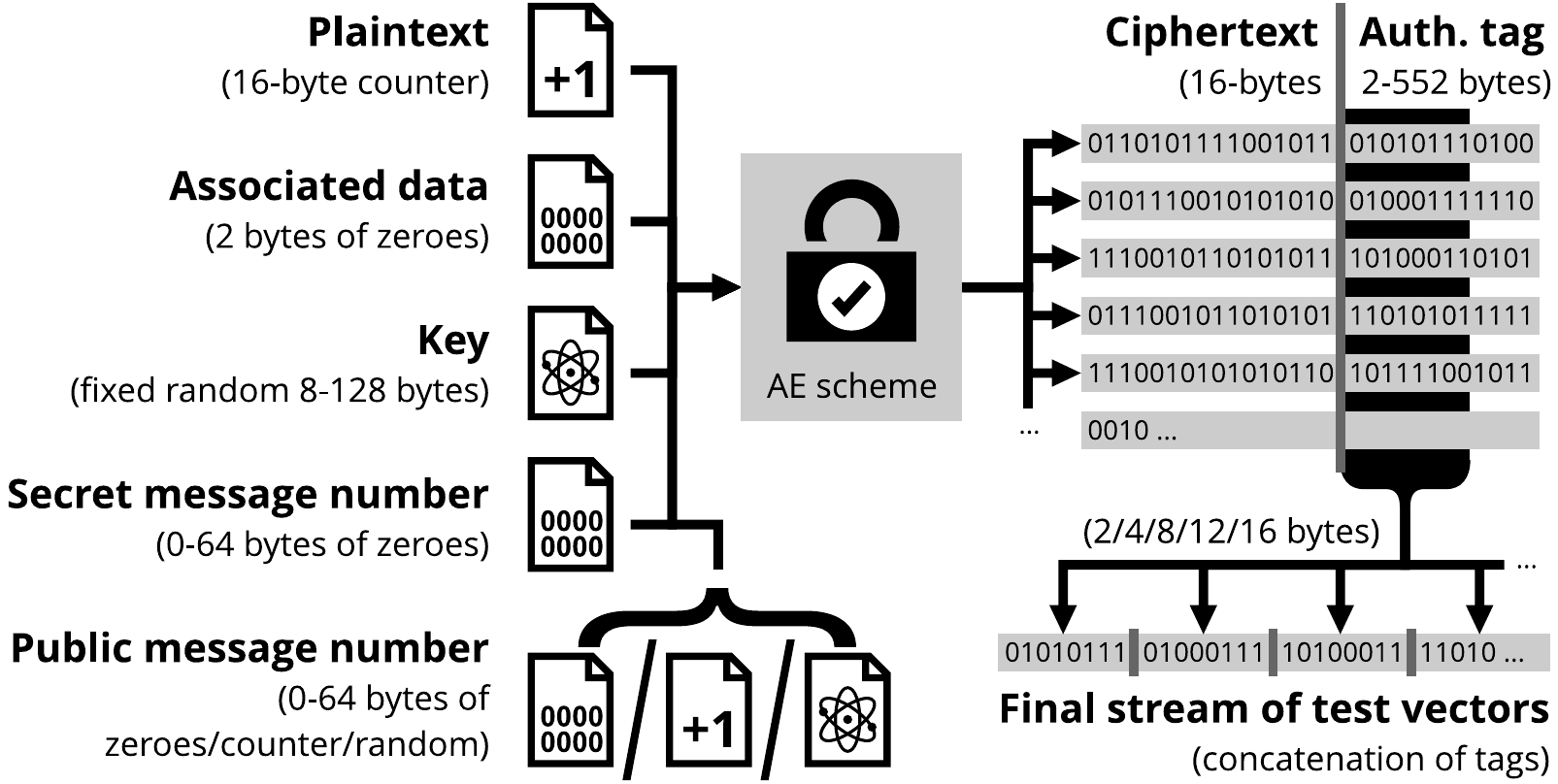}
\caption{The process of creating the tested data streams by individual \caesar{} candidates. The cipher is initialized as depicted in the diagram. The produced authentication tags are then concatenated to form a continuous stream suitable for randomness analysis.}
\label{fig:tv-generation}
\end{figure}

The aim of the performed experiments is to assess randomness of authentication tags produced by many authenticated encryption schemes. The same analysis could also be conducted on generated ciphertext, but that it out of the scope of this paper. In particular, we inspect tags provided by \caesar{} candidates in three independent scenarios differing in public message number setting. An overview of tag generation is given below and in Figure~\ref{fig:tv-generation}.

The cipher has five inputs: plaintext (encrypted and authenticated user input), associated data (authenticated user input), key, secret message number (secret nonce) and public message number (public nonce). The produced tag (extra ciphertext bytes when compared to the plaintext length) is determined by the cipher design. In the majority of the cases, this means 128 bits (16 bytes), but some candidates produce shorter tags (2, 4, 8 or 12 bytes). These tags were concatenated to form a continuous stream suitable for randomness assessment.

From the nature of the arguments prescribed by the \caesar{} API, the public message number is probably the argument to be most easily (unintentionally) misused. Security requirements for keys are well known, secret message numbers are usually not used, plaintext and associated data are mostly self-explanatory. Public message numbers are sometimes required to be unique (to have properties of nonces), but sometimes this is not necessary. In a way, we deem testing different modes of public message numbers as examining the robustness of the cipher design. The fields were initialized as follows:
\begin{itemize}
\item \textbf{Key}\\The key value was taken randomly but was fixed. For \eacirc{} (one of the used tools), 1\,000 independent runs used different keys to allow for variation (otherwise, the same numerical results would be produced).
\item \textbf{Associated data, secret message number}\\ We used two bytes of associated data; the length of the secret message number was determined by the cipher or the parameter set. Both fields' values were fixed to binary zeros. Note that only three ciphers used secret message numbers.
\item \textbf{Public message number}\\This was the only parameter explored in different settings:
\begin{itemize}
\item Fixed to a string of binary zeros for the whole time.
\item Increasing as a counter -- each value unique but similar to others.
\item Having each value completely random.
\end{itemize}
\item \textbf{Plaintext}\\The plaintext was 16 bytes long, formatted as a single counter starting from zero. We could not use fixed-value plaintext, because, in the case of fixed-value public message numbers, the produced tags would be identical (considering settings of the other arguments). A plaintext of binary zeros would have been possible in the other two modes for public message numbers, but we refrained from doing so to keep the experiments as comparable as possible (with as similar settings as possible).
\end{itemize}
In summary, if we denote the cipher as a function $F(plain, adata, key, smn, pmn)$ producing the authentication tag (the ciphertext is not used in our analysis, but inspecting it would also be interesting), the final analyzed stream in the scenario with random public message numbers looks as follows:
\begin{equation*}
\begin{split}
\textrm{Stream} =&\, F(0,0,rand_A,0,rand_1) \,||\, F(1,0,rand_A,0,rand_2) \,|| \\
&\, F(2,0,rand_A,0,rand_3) \,||\, F(3,0,rand_A,0,rand_4) \,||\, ...
\end{split}
\end{equation*}

\section{Randomness testing tools}
\label{sec:randomness-testing-tools}

The most common way of testing randomness is using statistical testing. From the multitude of available batteries, we used the following three: \niststs{} (older, yet still commonly used and a valid NIST standard), \dieharder{} (modern framework reimplementing other suites as well as adding brand new tests) and \testu{} (another modular framework implementing many tests).

Although the $p$-value of a randomness test focusing on a single characteristic has a clear statistical interpretation, the interpretation of results produced by testing suites is somewhat problematic. We need to determine what number of failed tests allows us to reject randomness of the assessed sequence while respecting the chosen significance level. For this, we use the methods proposed in 2015 by M.~S\'{y}s et al.~\cite{nist-sts-interpretation-syso}.

For all experiments, we chose the significance level of $\alpha = 1\%$, which is the default value for \niststs{}~\cite{nist-sts-documentation}. This setting keeps the type I.~error (false positives) reasonably low while preventing the type II.\ error (false negatives) to reach too high values.

We used \niststs{} version 2.1.1 with the default parameters (block lengths) for all tests. To comply with the minimal required stream length for individual tests~\cite{nist-sts-documentation}, we tested 100 independent 1\,000\,000 bit long sequences for each candidate. In summary, \niststs{} used about 12~MiB (about 700\,000 tags) of data from each candidate for each test.

\dieharder{} version 3.31.1 was used. The two parametrizable tests were configured with recommended values. The length of the input stream processed by \dieharder{} varies from test to test. The humblest (Diehard 3D-sphere test) required about 48~kiB, while the greediest one (Bit distribution test) took about 9.2~MiB. To ensure the best possible comparability with the other test suites, we again analyzed 100 independent samples of the input. In summary, \dieharder{} tests used between 4.7~MiB (about 300\,000 tags) and 916~MiB (about 60\,000\,000 tags) of input data for each candidate (depending on the test).

\testu{} was used in version 1.2.3. The most relevant sub-batteries are Rabbit, Alphabit and BlockAlphabit. These are intended for testing finite binary sequences. The length of the input stream taken by \testu{} can be set arbitrarily. To have an amount of data comparable to the other used batteries, we chose to process $2^{30}$ bits for each test. In summary, \testu{} thus used about 128~MiB (about 8\,400\,000 tags) of input data for each test.

EACirc represents an entirely different approach to testing data randomness: The main idea is to use supervised learning techniques based on evolutionary algorithms to design and further optimize a successful distinguisher -- a test determining whether its input comes from a truly random source or not. The distinguisher is represented as a hardware-like circuit consisting of simple interconnected functions. The used settings cause \eacirc{} to process approximately 2.24~MiB of data produced by the tested cryptoprimitive for a single \eacirc{} run. It amounts to about 2.24~GiB (about 150\,000\,000 tags) of data for a single experiment with 1\,000 runs.

\section{Results and interpretation}
\label{sec:results}

A selection of the numerical results can be seen in Table~\ref{tab:results}. The table aims for a representative selection of the interesting cases including all categories from the reference schemes to the algorithms that passed to the third (currently last) round. For the complete numerical results and exact reasoning, see~\cite{ukrop-mgr}.

\begin{sidewaystable}
\centering
\newcolumntype{C}{>{\centering\arraybackslash}X}
\newcolumntype{R}{>{\raggedleft\arraybackslash}X}
\begin{tabularx}{\textheight}{l@{}ll@{}|*{3}{|RRRR}}
& & & \multicolumn{4}{c|}{PMN fixed to zero} & \multicolumn{4}{c|}{PMN counter-based} & \multicolumn{4}{c}{PMN truly random} \\
\begin{tabular}[b]{@{}l}\large\textbf{Category}       \\ \scriptsize(\caesar{} round) \vspace{0.3em} \end{tabular} &
\begin{tabular}[b]{@{}l}\large\textbf{Cipher}       \\ \scriptsize(official name) \vspace{0.3em} \end{tabular} &
\begin{tabular}[b]{@{}l}\large\textbf{Candidate ID}       \\ \scriptsize(as used in SUPERCOP~\cite{supercop}) \vspace{0.3em}   \end{tabular} &
\rotatedHeader{\textbf{\eacirc{}}    \\ \scriptsize(proportion)} & \rotatedHeader{\textbf{\niststs{}} \\ \scriptsize(x/188)} &
\rotatedHeader{\textbf{\dieharder{}} \\ \scriptsize(x/55)}       & \rotatedHeader{\textbf{\testu{}}   \\ \scriptsize(x/159)} &
\rotatedHeader{\textbf{\eacirc{}}    \\ \scriptsize(proportion)} & \rotatedHeader{\textbf{\niststs{}} \\ \scriptsize(x/188)} &
\rotatedHeader{\textbf{\dieharder{}} \\ \scriptsize(x/55)}       & \rotatedHeader{\textbf{\testu{}}   \\ \scriptsize(x/159)} &
\rotatedHeader{\textbf{\eacirc{}}    \\ \scriptsize(proportion)} & \rotatedHeader{\textbf{\niststs{}} \\ \scriptsize(x/188)} &
\rotatedHeader{\textbf{\dieharder{}} \\ \scriptsize(x/55)}       & \rotatedHeader{\textbf{\testu{}}   \\ \scriptsize(x/159)} \\ \hline \hline
\multirow{2}{2cm}{\scriptsize Reference candidates}
& AES-GCM     & aes128gcmv1                & 1.000\cc & 23\cc & 1\cc & 11\cc & 0.014 & 187 & 52 & 157 & 0.019 & 187 & 52 & 158 \\ 
& AES-GCM     & aes256gcmv1                & 1.000\cc & 71\cc & 1\cc & 13\cc & 0.013 & 188 & 51 & 156 & 0.007 & 188 & 54 & 158 \\ 
\hline \hline
\multirow{2}{2cm}{\scriptsize Withdrawn candidates}
& Calico      & calicov8                   & 0.013 & 128\cc & 3\cc & 8\cc & 0.009 & 186 & 55 & 156 & 0.015 & 188 & 53 & 158 \\ 
\cline{2-15}
& Marble      & aes128marble4rv1           & 0.016 & 160\cc & 16\cc & 6\cc & 0.010 & 168\cc & 14\cc & 8\cc & 0.010 & 160\cc & 16\cc & 6\cc \\ 
\hline \hline
\multirow{8}{2cm}{\scriptsize First-round candidates}
& AES-CMCC    & cmcc22v1                   & 0.008 & 53\cc & 1\cc & 4\cc & 0.011 & 46\cc & 2\cc & 8\cc & 0.008 & 187 & 54 & 156 \\ 
& AES-CMCC    & cmcc24v1                   & 0.005 & 43\cc & 3\cc & 3\cc & 0.008 & 188 & 50 & 155 & 0.023 & 186 & 50 & 152 \\ 
& AES-CMCC    & cmcc42v1                   & 0.011 & 87\cc & 2\cc & 3\cc & 0.008 & 86\cc & 2\cc & 4\cc & 0.015 & 182 & 54 & 154 \\ 
& AES-CMCC    & cmcc44v1                   & 0.008 & 85\cc & 4\cc & 5\cc & 0.013 & 183 & 52 & 155 & 0.007 & 188 & 53 & 152 \\ 
& AES-CMCC    & cmcc84v1                   & 0.014 & 147\cc & 7\cc & 4\cc & 0.009 & 184 & 53 & 156 & 0.007 & 182 & 48\cc & 158 \\ 
\cline{2-15}
& Enchilada   & enchilada128v1             & 1.000\cc & 71\cc & 2\cc & 15\cc & 0.017 & 187 & 53 & 157 & 0.010 & 186 & 52 & 155 \\ 
& Enchilada   & enchilada256v1             & 1.000\cc & 77\cc & 1\cc & 11\cc & 0.013 & 188 & 54 & 156 & 0.016 & 188 & 53 & 155 \\ 
\cline{2-15}
& Raviyoyla   & raviyoylav1                & 1.000\cc & 22\cc & 2\cc & 7\cc & 1.000\cc & 148\cc & 28\cc & 24\cc & 0.295\cc & 186 & 51 & 144 \\ 
\hline \hline
\multirow{2}{2cm}{\scriptsize Second-round candidates}
& TriviA-ck   & trivia0v1                  & 0.999\cc & 140\cc & 5\cc & 8\cc & 0.015 & 186 & 52 & 157 & 0.005 & 187 & 55 & 154 \\ 
& TriviA-ck   & trivia128v1                & 0.993\cc & 158\cc & 12\cc & 8\cc & 0.017 & 188 & 53 & 158 & 0.009 & 188 & 54 & 157 \\ 
\hline \hline
\multirow{2}{2cm}{\scriptsize Third-round candidates}
& AEZ         & aezv1                      & 0.014 & 169\cc & 15\cc & 9\cc & 0.015 & 187 & 52 & 155 & 0.010 & 188 & 52 & 157 \\ 
& AEZ         & aezv3                      & 0.016 & 164\cc & 13\cc & 6\cc & 0.011 & 188 & 53 & 157 & 0.009 & 185 & 50 & 156 \\ 
\end{tabularx}
\caption{The selection of ciphers with interesting results from different categories (from reference schemes to 3\textsuperscript{rd} round candidates). The numbers in columns of statistical batteries represent the number of passed tests (should be close to all for random stream), while \eacirc{} displays the ratio of runs rejecting randomness (should be around 0.010 for random stream). For the ease of comprehension, if the result rejects randomness of the particular stream, the cell is gray-colored. Note that this is only a subset of the tested candidates (most of the omitted ones have results similar to those of \textit{AEZ}). For complete numerical results and threshold values, see~\cite{ukrop-mgr}.}
\label{tab:results}
\end{sidewaystable}

Firstly, let us compare the outcomes for the three inspected public message number modes. We expected the random-valued to perform the best, followed by counter-based and then by zero-fixed public message numbers. We reasoned that the more differences there will be among the used values, the easier it will be for the cipher to produce a random-looking tag (since it has more entropy to start from). As stated in the submission call, the ciphers were allowed to lose all security in case of reused (public message number, private message number)-pair under the same key. Nevertheless, we expected some (albeit not many) ciphers will be able to retain the apparent randomness of the produced tag -- even though it would require an adamant avalanche effect (all arguments are identical apart from a few bits in plaintext).

From the conducted experiments we see that the primary hypothesis (random values performing better than a counter and much better than zeros) was confirmed. However, none out of the tested candidates passed with the public message numbers set to zero. The single bit change in plaintext with all other arguments fixed might not have been enough to cause the avalanche effect needed to produce a tag looking sufficiently random.

Secondly, let us inspect the results for the individual candidates (see Table~\ref{tab:results}). Tags of just five ciphers (\textit{AES/GCM}, \textit{Marble}, \textit{AEC-CMCC}, \textit{AES-CPFB}, \textit{Raviyoyla}) were distinguishable from random streams with counter-valued public message numbers. Three of these ciphers (\textit{Marble}, \textit{AES-CMCC}, \textit{AES-CPFB}) also failed in the random-valued scenario. The evidence is still too weak to deem the designs insecure -- it may merely be the case they produce a constant delimiter between the ciphertext and tag, violating the statistical randomness of the created tag. To draw any conclusions, a detailed inspection of the ciphers would need to be performed. It is, however, worth mentioning that no candidates failing in either counter- or random-valued scenario were selected by the \caesar{} committee to the second round of the competition.

Apart from the findings for the \caesar{} candidates, the results allow us to gain insights into the capabilities of the used randomness testing tools. Based on the previous works~\cite{secrypt2014}, we expected the randomness distinguishing abilities of \eacirc{} and \niststs{} will be similar while both will be surpassed by \dieharder{} and \testu{}. On the one hand, the observed results showed many deficiencies of \eacirc{} -- it performed worse than \niststs{} in given tested scenarios. On the other, all three statistical batteries achieved comparable results. However, before any conclusions on the quality of the batteries are drawn, one has to be aware there are many domains in which these tools remain incomparable. They inspect different amounts of data and have different modes of operation (batteries see the stream as a whole, \eacirc{} processes short, distinct test vectors).

There is one case contrary to the general behavior observed above (see Table~\ref{tab:results}): \textit{Raviyoyla} with randomly initialized public message numbers for each test vector seems to be successfully rejected from the random stream by \eacirc{} although none of the statistical batteries support such result. It appears very promising but also requires additional inspection and enhanced testing to announce a case of \eacirc{} surpassing all tested statistical batteries.

The results lead us to several interesting hypotheses requiring further investigation. The candidates failing in randomness tests would deserve a deeper manual inspection to prove their potential (in)security. The used statistical testing suites themselves would be an interesting target for further research. It turned out that interpretation of test suites is quite difficult, and thorough research on test interdependence is necessary. Another perspective direction would be weakening the cipher designs (e.g.\ by limiting the number of internal rounds) to achieve a fine-grained comparison of the used tools.

\section{Summary}

We have set off to examine modern authenticated encryption systems from the point of resistance against common developer misconfiguration. In the end, we assessed outputs from 168 distinct schemes (all but six \caesar{} submissions) in three different configurations using multiple software tools (\niststs{}, \dieharder{}, \testu{} and \eacirc{}).

We examined a scenario with random (but fixed) keys, counter-based plaintext and three different settings of public message numbers. As expected, tags produced in settings with random public message numbers fared better than the ones from counter-based configurations. Both did better than tags from fixed-value public message numbers -- no submission had an avalanche effect strong enough to produce random-looking tags in the scenario where all test vectors had the same public message numbers.

Only three \caesar{} submissions (\textit{Marble}, \textit{AEC-CMCC}, \textit{Raviyoyla}) failed to produce seemingly random tags with counter-based public message numbers. For entirely random public message numbers, only \textit{Marble} failed convincingly. \textit{AEC-CMCC} achieved a borderline value in \dieharder{} and passed in other tools. \textit{Raviyoyla} seems to have failed according to \eacirc{} -- this case is suspicious and worth of further investigation since it is the only case where \eacirc{} surpassed the other tools. Importantly, none of these candidates made it to the second round of the competition (indirectly supporting our results).

Regarding the tools used for tag evaluation, \eacirc{} seemed to be the least suitable for the given task, being beaten by all the statistical batteries. The batteries themselves (\niststs{}, \dieharder{} and \testu{}) produced comparable results. The only exception is the case of \textit{Raviyoyla}, in which \eacirc{} seems to have outperformed all the other tools. However, when making comparisons, one has to take into account the amount of data inspected by each tool and their different modes of operation.

All in all, not even the state-of-the-art authenticated encryption designs do not have avalanche effect strong enough in the case with zero-fixed public message numbers. Although not forming a direct practical attack on the ciphers, it breaks the semantic security of the scheme, since the attacker can distinguish two messages based on the leakage present in inspected scenarios. A security-optimist from the introduction may see this as an interesting area for further improvements.

\subsubsection*{Acknowledgments}
We acknowledge the support of Czech Science Foundation, project GA16-08565S.
Access to computing and storage facilities owned by parties and projects contributing to the National Grid Infrastructure MetaCentrum provided under the program “Projects of Projects of Large Research, Development, and Innovations Infrastructures” (CESNET LM2015042), is greatly appreciated.

\bibliographystyle{eptcs}
\bibliography{memics}
\end{document}